\documentclass{iau307}
\usepackage{graphicx}
\usepackage{natbib}
\usepackage{url}
\usepackage{dtklogos}
\bibpunct{(}{)}{;}{a}{}{,}

\title[Tidal interactions in rotating multiple stars]
{Tidal interactions in rotating multiple stars and their impact on their evolution}

\author[P. Auclair-Desrotour, S. Mathis, C. Le Poncin-Lafitte]
{P. Auclair-Desrotour$^{1}$,
S. Mathis$^{2}$ and C. Le Poncin-Lafitte$^{3}$}

\affiliation{
$^1$IMCCE, Observatoire de Paris, UMR 8028 du CNRS, UPMC,\\ 77 Av. Denfert-Rochereau, 75014 Paris, France\\
$^2$Laboratoire AIM Paris-Saclay, CEA/DSM - CNRS - Universit\'e Paris Diderot, IRFU/SAp Centre de Saclay, F-91191 Gif-sur-Yvette Cedex, France\\
$^3$SYRTE, Observatoire de Paris, UMR 8630 du CNRS, UPMC,\\ 77 Av. Denfert-Rochereau, 75014 Paris, France\\
email: {\tt pierre.auclair-desrotour@obspm.fr, stephane.mathis@cea.fr, christophe.leponcin@obspm.fr}}

\pubyear{2014}
\volume{307} 
\pagerange{}
\jname{New windows on massive stars: asteroseismology, interferometry, and spectropolarimetry}
\editors{G. Meynet, C. Georgy, J.H. Groh \& Ph. Stee, eds.}

\begin{document}

\maketitle

\begin{abstract}
Tidal dissipation in stars is one of the key physical mechanisms that drive the evolution of binary and multiple stars. As in the Earth oceans, it corresponds to the resonant excitation of their eigenmodes of oscillation and their damping. Therefore, it strongly depends on the internal structure, rotation, and dissipative mechanisms in each component. In this work, we present a local analytical modeling of tidal gravito-inertial waves excited in stellar convective and radiative regions respectively. This model allows us to understand in details the properties of the resonant tidal dissipation as a function of the excitation frequencies, the rotation, the stratification, and the viscous and thermal properties of the studied fluid regions. Then, the frequencies, height, width at half-height, and number of resonances as well as the non-resonant equilibrium tide are derived analytically in asymptotic regimes that are relevant in stellar interiors. Finally, we demonstrate how viscous dissipation of tidal waves leads to a strongly erratic orbital evolution in the case of a coplanar binary system. We characterize such a non-regular dynamics as a function of the height and width of resonances, which have been previously characterized thanks to our local fluid model.
\end{abstract}

Many stars of our galaxy are components of close binary or multiple stellar systems. In such multiple stars, tides play a key role to modify the rotational evolution of the components and of their orbit \citep[][Langer in this volume]{deminketal2013}. The same applies to the case of star-planet systems \citep{Albrechtetal2012}. In this framework, \cite{Zahn1977} and \cite{OL2007} demonstrated that this the dissipation of the kinetic energy of the low-frequency eigenmodes of oscillation of the stars excited by tides (i.e. tidal inertial waves driven by the Coriolis acceleration in convection zones and gravito-inertial waves driven by the buoyancy restoring force and the Coriolis acceleration in stably stratified radiation zones) that drive tidal migration, the circularization of the orbits, and the synchronization and alignment of the spins \citep[][]{BO2009}.\\

\begin{figure}[h!]
\centering
\includegraphics[width=0.25\textwidth]{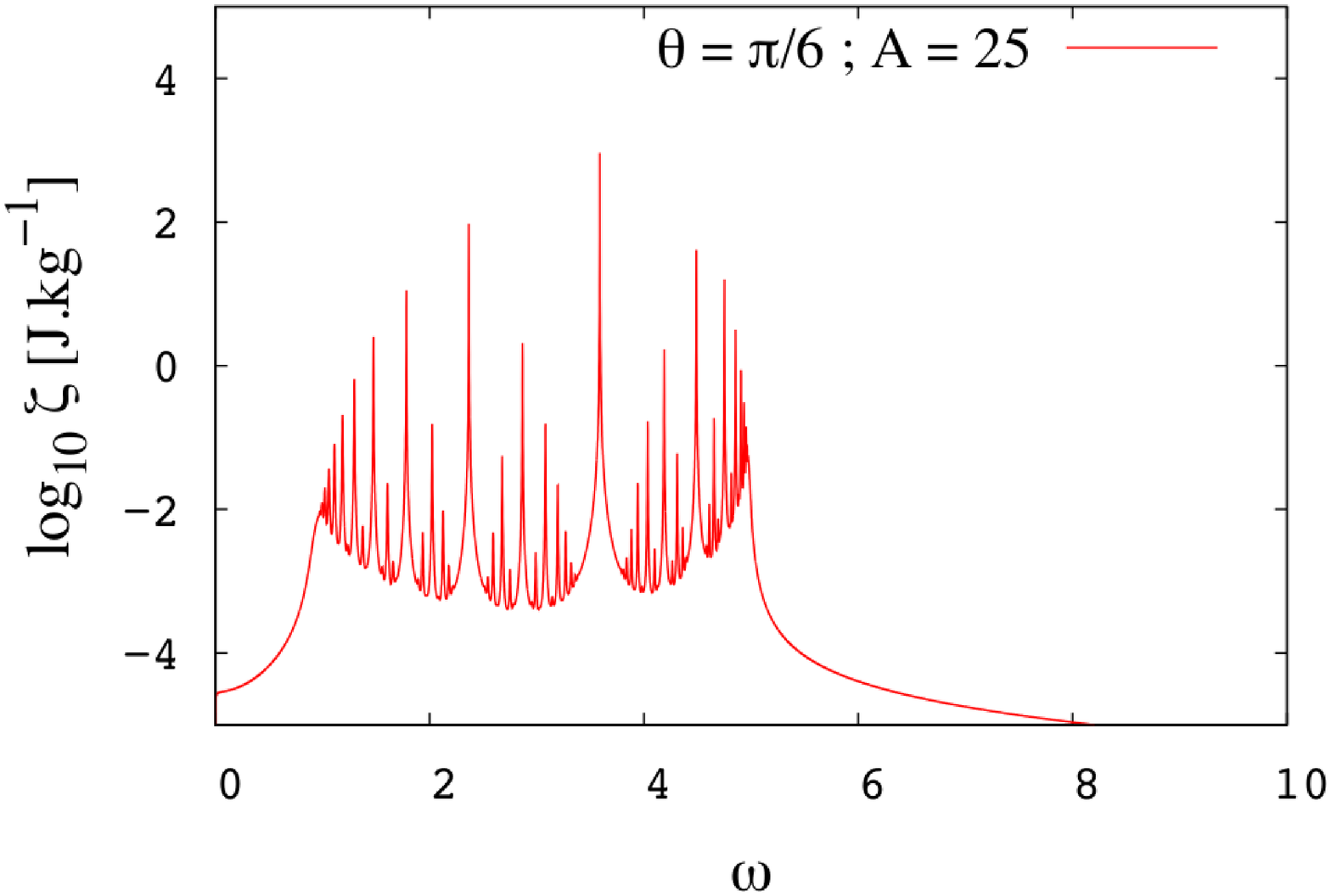}
\includegraphics[width=0.18\textwidth]{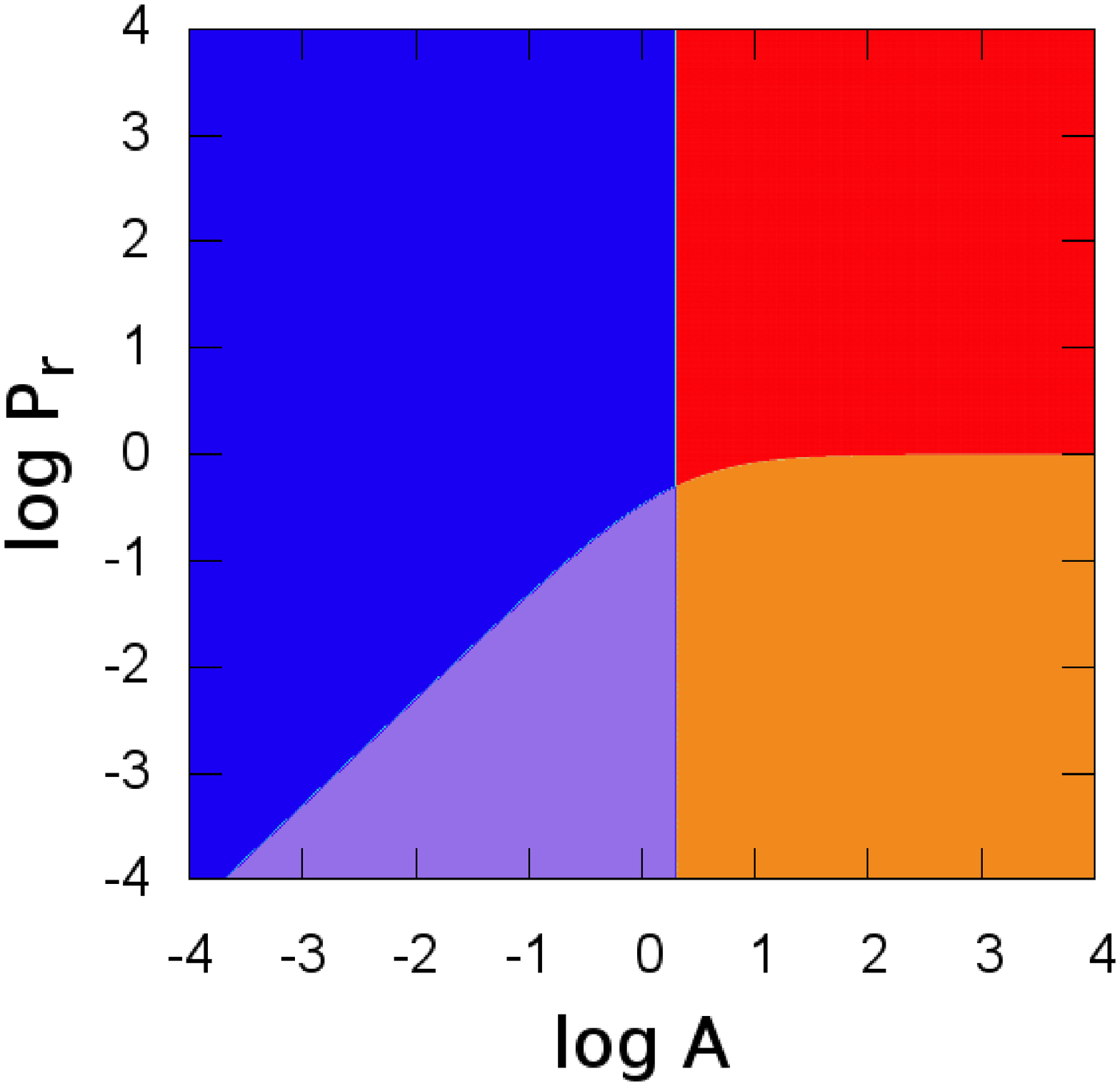} 
\includegraphics[width=0.25\textwidth]{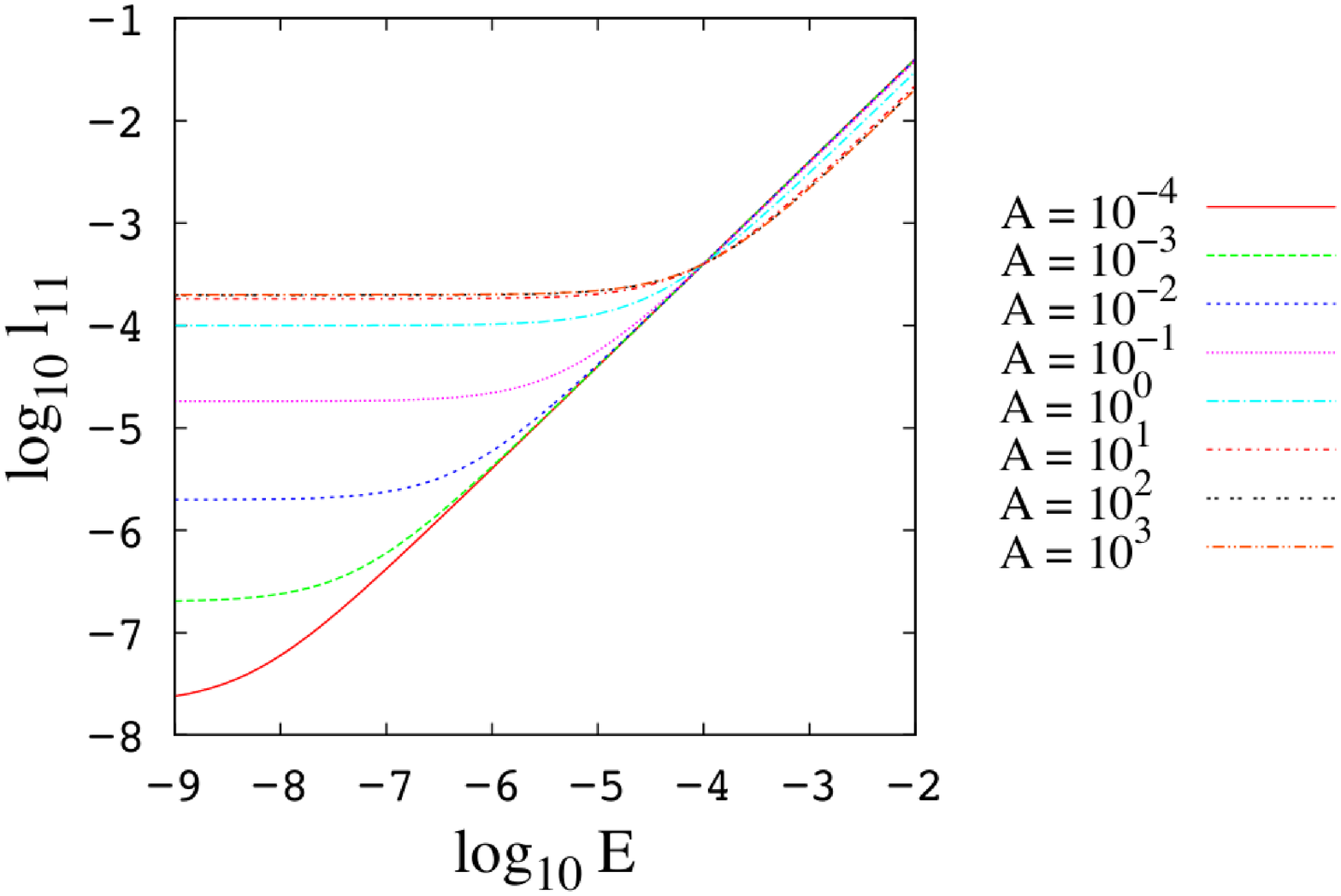} 
\caption{{\bf Left:} Example of the frequency-spectra of the viscous dissipation of tidal gravito-inertial waves. {\bf Middle:} the four asymptotic regimes for tidal waves in stars. {\bf Right:} Example of scaling law for the width at mid-height of the resonance $\left\{m,n\right\}\equiv\left\{1,1\right\}$ as a function of $E$ and $A$.}
\label{fig1}
\end{figure}

\section*{Tidal waves in a box: a reduced model to understand tidal dissipation in stars}
In this study, we consider a two-body system. The central body A, which has a mass $M_{\rm A}$ and a radius $R_{\rm A}$, rotates with an angular velocity $\Omega$. The punctual-mass tidal perturber B ($M_{\rm B}$) has an orbit relative to A with a semi-major axis $a$ and a mean motion ${\widetilde n}$. We compute the viscous dissipation of tidal gravito-inertial waves, which are driven by the buoyancy and the Coriolis acceleration, using a local Cartesian model that generalizes the one of \cite{OL2004}. We thus consider a box of characteristic length $L$ inclined with a co-latitude $\theta$ relatively to the rotation axis. The fluid has a density $\rho$, a kinematic viscosity $\nu$, a thermal diffusivity $\kappa$, and a Brunt-V\"ais\"al\"a frequency $N$ (with $N\approx0$ in convective regions). The control parameters of the system are: i) the Froude number $A=\left[N/\left(2 \Omega\right)\right]^2$ that compares the buoyancy force and the Coriolis acceleration, ii) the Ekman number $E=\left(2 \pi^2 \nu\right)/\left(\Omega L^2\right)$ that compares the viscous diffusion time to $t_{\Omega}=\left(2\Omega\right)^{-1}$, and iii) the thermal number $K=\left(2 \pi^2 \kappa\right)/\left(\Omega L^2\right)$ that compare the thermal diffusion time to $t_{\Omega}$. The viscous dissipation (fig. \ref{fig1}, left) strongly depends on the tidal frequency $ \sigma = 2 \left({\widetilde n} - \Omega \right)  $ and of the fluid parameters ($\Omega$, $A$, $E$, $K$). This complex behavior must be taken into account when studying the dynamical evolution of the system \citep{MLP09}. Four asymptotic regimes (see fig. \ref{fig1}, middle) are identified. In convection zones ($A\le0$), we identify tidal inertial waves dissipated by viscous (in blue; $P_{\rm r}=\nu/\kappa$ is the Prandt number) or thermal (in purple) diffusions. Respectively, in radiation zones ($A>0$), we identify tidal gravito-inertial waves dissipated by thermal (in orange) or viscous (in red) diffusions. In these regimes, we have derived analytic scaling laws for the eigenfrequencies $\left(\omega_{mn};m\hbox{ and }n\hbox{ correspond to the horizontal and vertical wave numbers}\right)$ of resonances, their width at half-height $\left(l_{mn}\right)$ (fig. \ref{fig1}, right), their height $\left(h_{mn}\right)$, their number $\left(N_{\rm kc}\right)$, and the amplitude of the non-resonant background ($H_{\rm bg}$) that corresponds to the so-called equilibrium tide \citep[e.g.][]{RMZ2012}. Finally, we demonstrated that in the case of such resonant tidal dissipation spectra, the orbital evolution becomes erratic \citep{WS1999, ADLPM2014}. This behavior is observed through jumps in semi-major axis that scale as $\Delta a/a \propto l H^{1/4}$, where the width at mid-height $ l $ and height $ H $ of resonances are characterized thanks to obtained scaling laws.

\bibliographystyle{iau307}
\bibliography{Biblio_Mathis_proc3}

\end{document}